\documentclass[superscriptaddress,aps,prl,twocolumn,showpacs,preprintnumbers,amsmath,amssymb]{revtex4}
\usepackage{times}
\usepackage{graphicx}
\usepackage{dcolumn}
\usepackage{bm}

\newcommand{\um}{$~\mu$m}
\newcommand{\nm}{~nm}

\newcommand{\LX}{$L_X$}
\newcommand{\LY}{$L_Y$}

\begin{document}
\title{Highly Confined Hybrid Spoof Surface Plasmons in Ultra-thin
Metal/Dielectric Heterostructures}

\author{S. Hossein Mousavi}
\affiliation{Department of Physics, University of Texas at Austin,
Austin, Texas 78712}
\author{Alexander B. Khanikaev}
\affiliation{Department of Physics, University of Texas at Austin,
Austin, Texas 78712}
\author{Burton Neuner III}
\affiliation{Department of Physics, University of Texas at Austin,
Austin, Texas 78712}
\author{Yoav Avitzour}
\affiliation{Department of Physics, University of Texas at Austin,
Austin, Texas 78712}
\author{Dmitriy Korobkin}
\affiliation{Department of Physics, University of Texas at Austin,
Austin, Texas 78712}
\author{Gabriel Ferro}
\affiliation{Laboratoire des Multimat{\'e}riaux et Interfaces,
  Universit\'{e} Claude Bernard Lyon 1, 69622 Villeurbanne, France}
\author{Gennady Shvets}
\email{gena@physics.utexas.edu}
\affiliation{Department of Physics, University of Texas at Austin,
Austin, Texas 78712}
\date{\today}
\begin{abstract}

Highly confined ``spoof'' surface plasmon-like (SSP) modes are
theoretically predicted to exist in a perforated metal film coated
with a thin dielectric layer. Strong modes confinement
results from the additional waveguiding by the layer. Spectral
characteristics, field distribution, and lifetime of these SSPs
are tunable by the holes' size and shape. SSPs exist both
above and below the light line, offering two classes of
applications: ``perfect'' far-field absorption and to efficient
emission into guided modes. It is experimentally shown that these
plasmon-like modes can turn thin, weakly-absorbing semiconductor
films into perfect absorbers.
\end{abstract}

\maketitle
{\em Introduction.} The surface plasmon (SP) is one of the linchpins of the field of
sub-diffraction optics~\cite{barnes_nature03}: by penetrating
below the metal surface~\cite{Raether_book}, electromagnetic waves 
can be localized beyond diffraction limit. This
opens a wide range of applications in spectroscopy, photonic
circuits, solar cells, and other technologically important
areas~\cite{halas_natphot07,Atwater,bozh_nature06}. For longer
wavelengths, metals begin resembling perfect electric conductors
(PECs), wave penetration into the metal becomes negligible, and
other approaches to light localization and concentration must be
found. It has been recently
discovered~\cite{Pendry_sc,Abajo_PRL,Abajo_OptExp,Abajo_Review,Maier}
that SPs can be mimicked by perforating the metal surface with an
array of sub-wavelength holes. Dispersion and confinement of the
resulting ``spoof'' surface plasmons (SSPs) are defined (and can
be tuned) by the holes' size and geometry. Thus, the SSP-based
approach to light confinement is crucial for bringing the
advantages of plasmonics into the longer-wavelength spectral
range, which is especially important for various infrared
applications, including surface-enhanced infrared absorption
(SEIRA) spectroscopy ~\cite{Ozawa_JPC91,Halas_CPL08}, on-chip
light sources~\cite{brongersma_optexp08}, and infrared
detectors~\cite{painter_optexp10,shawnlin_NL10}.
\begin{figure}[ht]
    \centering
    \includegraphics[clip=true,scale=0.10]{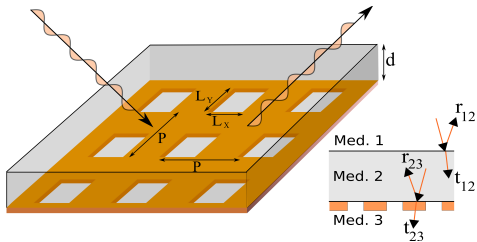}
    \caption{(color online). (Left) Schematic for the detection of hybrid SSPs
      supported by a heterostructure comprised of a perforated metal film
      and a dielectric film. (Right) Side view and definitions of
      the reflection/transmission coefficients used in the
      text.}\label{Fig:ACM}
\end{figure}

A major drawback of SSPs supported by metal films with simple
perforation geometries (e.g., circular or rectangular holes) is that they are
very weakly confined for most frequencies and, therefore, are not
suitable for the above applications. More sophisticated geometries
that have been shown~\cite{Maier,Sambles_PRL} to improve
confinement in the microwave spectral range are not practical for
optical applications. In this Letter, we resolve the problem of
poor confinement of conventional SSPs while retaining their
important advantage: the ability to control SSP's frequency and refractive index using variable size and shape of the metal holes. This is achieved by combining metal structuring with
the traditional method of electromagnetic field confinement,
waveguiding. We theoretically predict and experimentally confirm
that a new class of SSPs exists in the ultrathin heterostructure
comprised of a slab of high-index dielectric material and a
perforated metal film (Fig.~\ref{Fig:ACM}). The resulting SSPs owe
their unusual electromagnetic properties to their hybrid nature:
the conventional SSP modes inherent to any periodic metal
structure are hybridized with conventional modes of dielectric
films: guided waves (GW) and leaky Fabry-Per\'{o}t (FP) modes. The
hybrid SSPs represent an important improvement on both
conventional guided modes and SSPs: they are better confined than
either one of the two (Fig.~\ref{Fig:FieldProfiles}), while their spectral properties and lifetimes
are controllable by the size and shape of the metallic perforations. As
a specific example and possible application of the proposed
concept, we demonstrate that hybrid SSPs can turn low-absorbing
semiconductor (specifically, SiC) films into near-perfect
absorbers of infrared radiation. The choice of SiC owes to the
strong dependence of its dielectric permittivity $\epsilon_{\rm
SiC}(\omega)$$\equiv$$n^2(\omega)$ on frequency
$\omega$$\equiv$$2\pi c/\lambda$: $\epsilon_{\rm SiC}$ changes by
a factor $2$ as $\lambda$ changes by $8\%$~\cite{neuner_jpc10}.


\begin{figure}[ht]\centering
  \includegraphics[clip=true,scale=0.45]{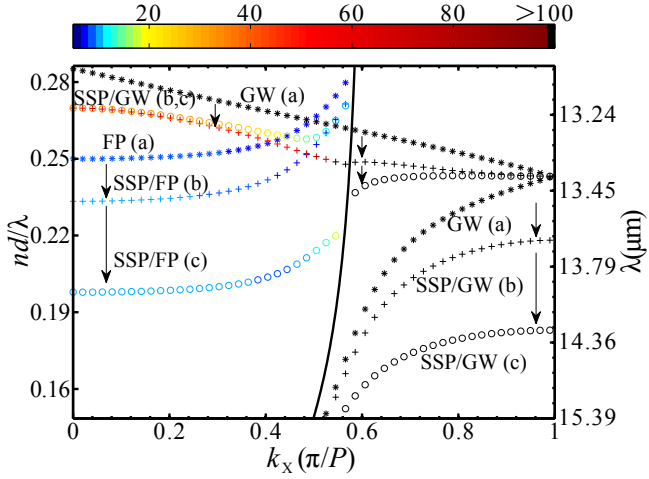}
  \caption{(color online). Family of dispersion curves of the {\it
      p}-polarized eigenmodes of the heterostructure shown in
    Fig.~\ref{Fig:ACM} for different hole sizes: stars (a) - unperforated
    metal; pluses (b) - medium size holes (\LX=\LY=1.2\um{});
    circles (c) - large holes (\LX=\LY=1.8\um{}). Periodicity $P$=$3.84$\um, SiC film thickness $d$=$570$\nm.
    Color: the quality factor of the modes. Solid curve is the air light line.}
  \label{Fig:DD}
\end{figure}

{\em Theoretical description.} It is instructive to start the theoretical description of hybrid
SSPs by considering the eigenmodes (both leaky and fully-confined) of the
{\it unperforated} heterostructure.
Mathematically, this is done by recalling the
standard Airy expression~\cite{born_and_wolf} for the reflectivity coefficient of the heterostructure
$r_{slab}$:
\begin{equation}
r_{\rm slab}=r_{12}+t_{12}t_{21}\;exp(2i\delta)\;r_{23}\;
Z,\;\;\;\;\delta=k_zd,
\label{Eq:refl_slab}
\end{equation}
where the reflection (transmission) coefficients
$r_{ij}\equiv|r_{ij}|e^{i\phi_{ij}}$ ($t_{ij}$) at the
$i$-th/$j$-th medium interface are illustrated by
Fig.~\ref{Fig:ACM}, $Z \equiv 1/(1-r_{21}r_{23}e^{2i\delta})$
accounts for multiple reflections, and $k_z$ is the propagation
wave number normal to the interface. All eigenmodes of the
unperforated heterostructure (including leaky and fully-guided
modes) correspond to the poles of $r_{\rm slab}$. These poles are given by $Z^{-1}$=$0$, giving rise to
 the following necessary condition that must be satisfied by any eigenmode:
\begin{equation}
  \label{Eq:ReqPhase}
  \phi_{21}+\phi_{23}+2\,Re\{k_z\}d =2\pi{}l,
\end{equation}
where $l$ is an integer number. Thin dielectric films with $d <
\lambda/2n(\lambda)$ correspond to $l$=$1$, which will be assumed
here. Note that $\phi_{23}$=$\pi$ in the case of unperforated
PEC film, and the phase
$\phi_{21}$ of the Fresnel's reflection coefficient satisfies
$\phi_{21}$=$0$ above the light line, but can depart from zero
below the light line. The frequencies of the two eigenmodes
found by solving $r_{slab}^{-1}=0$ for real wavenumbers $\vec{k}_{||}$ along
the interface correspond to the two eigenmodes: (i) the leaky FP mode above the light line, and (ii) the confined GW mode below the light line. 
For a specific case of a SiC film with thickness
$d=570$\nm, the dispersion relations $\omega(\vec{k}_{||})\equiv
\omega_r(\vec{k}_{||}) + i \omega_{i}(\vec{k}_{||})$ for these two eigenmodes are plotted in
Fig.~\ref{Fig:DD}. For $k_{||}=0$, the FP mode satisfies the quarter-wavelength condition $d$=$\lambda_{\rm FP}/4n(\lambda_{\rm FP})$. Because the FP mode is leaky, it has a low
quality factor $Q=\omega_r/\omega_i$. However, coupling to the GW mode requires a high-index prism or a diffraction grating,
and its $Q$ is limited only by the material losses. As shown
below, when the metal film is patterned, both FP and GW modes can
hybridize with SSPs giving rise to highly confined modes whose
spectral properties can be tuned by the size and shape of the
perforations, and whose confinement drastically exceeds that of
either the SSPs or the modes of the unperforated structure (GW and FP modes).

Next, we consider a periodically perforated metal film as shown in
Fig.~\ref{Fig:ACM}. SSPs are introduced by generalizing
Eq.~(\ref{Eq:refl_slab}) to include the effects of diffraction on
the periodic hole array. Specifically, using the standard
scattering matrix formulation, the reflection coefficient, which connects different diffractive
orders, is expressed in the matrix form:
\begin{equation}
\hat{r}_{\rm
slab}=\hat{r}_{12}+\hat{t}_{21}\;exp(i\hat{\delta})\;\hat{r}_{23}\;
\hat{Z}\;exp(i\hat{\delta})\;\hat{t}_{12},
\label{Eq:refl_slab_diffraction}
\end{equation}
where $\hat{Z}$ is the multi-pass matrix
given by $\hat{Z}=(\hat{I}-exp(i\hat{\delta})\;
\hat{r}_{21}exp(i\hat{\delta})\;\hat{r}_{23})^{-1}$. Because of the
sub-wavelength periodicity of the hole array,
the diagonal matrix element $r_{\rm slab}$=$(\hat{r}_{\rm slab})_{00}$ expresses the
experimentally-accessible reflection coefficient. Complex
eigenfrequencies of the hybrid SSPs correspond to the roots of the
equation ${\displaystyle Det}\,[\hat{r}^{-1}_{\rm slab}
(\omega,\vec{k})]$=$0$.

\begin{figure}[ht]
    \centering
    \includegraphics[scale=0.25]{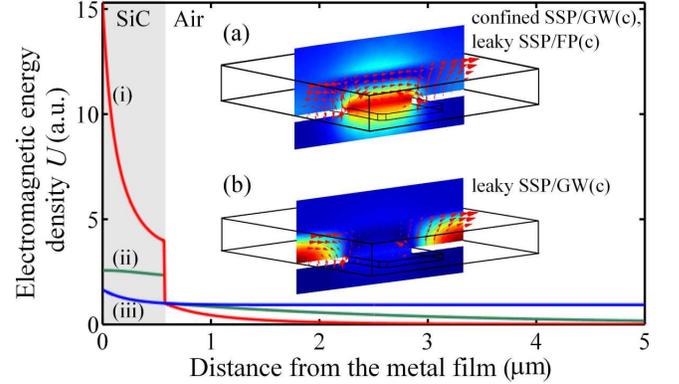}
    \caption{(color online). Transverse distribution of electromagnetic energy density averagesd over the unit cell for the case of (i) SSP/GW of perforated heterostructure, (ii) GW of unperforated heterostructure, and (iii) SSP of uncoated perforated PEC. Parameters are the same as in Fig.~\ref{Fig:DD}. Insets show the fields corresponding to the excitation of hybrid SSPs given in Fig.~\ref{Fig:DD}.}
    \label{Fig:FieldProfiles}
\end{figure}

To make a further connection of hybrid SSPs to the modes of
unperforated heterostructure, we introduce two effective
(tilded) complex quantities for a perforated film:
$\tilde{r}_{23}{\tilde{Z}} \equiv (\hat{r}_{23}\hat{Z})_{00}$ and
$\tilde{Z} \equiv 1/(1-r_{21}\tilde{r}_{23}e^{2i\delta})$. By
comparing Eq.~(\ref{Eq:refl_slab}) to Eq.~(\ref{Eq:refl_slab_diffraction}) and by using the above definitions, we
find that
\begin{equation}
\tilde{r}_{23}\equiv |\tilde{r}_{23}| e^{i\tilde{\phi}_{23}}=\frac{\sum}{1+r_{21}\:e^{2i\delta}\:\sum},
\label{Eq:r23}
\end{equation} where
$\sum\equiv\sum_{\vec{G}}(\hat{r}_{23})_{0\vec{G}}\;
(\hat{Z})_{\vec{G}0}$ involves summation over all the reciprocal lattice vectors $\vec{G}$ (i.e., all the diffracted
orders). These
intuitive definitions of the effective quantities enable us to use
the scalar Eqs.~(\ref{Eq:refl_slab}, \ref{Eq:ReqPhase}) by
replacing transmission/reflection coefficients and the
corresponding phases with their effective counterparts. For
example, the modified resonant phase-matching condition expressed
by Eq.~(\ref{Eq:ReqPhase}) is obtained by replacing
$\phi_{23}$=$\pi$ with $\tilde{\phi}_{23}$, which can considerably
deviate from $\pi$. As will be shown below, the size and shape of
the hole can strongly influence $\tilde{\phi}_{23}$, thereby
shifting the absorption peak from the Fabry-Per\'{o}t resonance
$d$=$\lambda_{\rm FP}/4n(\lambda_{\rm FP})$.

Assuming a square array of holes in the PEC metal with the
sub-diffraction period $P=3.84$~$\mu$m and a thin SiC film with
$d=570$\nm, we calculated the scattering matrices
($\hat{r}_{23}$ and $\hat{t}_{23}$) using the semi-analytic modal
matching technique~\cite{Martin-Moreno_PRL}, and analytically
continued it into the complex $\omega$-plane. The resulting
dispersion curves for hybrid SSPs both above (leaky SSPs) and
below (confined SSPs) the light line are plotted in
Fig.~\ref{Fig:DD}, where $j$=(a,b,c) labels the hole size (see
 caption), and the color indicates the quality factor $Q$ of the
modes.

First we focus on dispersion curves below the light line: the
guided mode of the unperforated structure GW(a), and the two
hybrid SSPs (SSP/GW(b) and SSP/GW(c)) whose dispersion curves are
controlled by the square holes' size. As the hole size $L_X=L_Y$
increases, the dispersion curves depart from the light line,
thereby indicating high spatial confinement. Flattening of the
dispersion curves for larger hole sizes also indicates an
increasing density of states which can be utilized for
plasmon-enhanced infrared emission~\cite{brongersma_optexp08}.
Strong concentration of the magnetic field of the hybrid SSP/GW
mode inside the hole (Fig.~\ref{Fig:FieldProfiles} inset (a))
explains the high sensitivity of this mode to the size and shape
of the hole. Figure~\ref{Fig:FieldProfiles} illustrates that the confinement of the
SSP/GWs exceeds that of both classes of waves (GWs and
pure SSPs without the dielectric film) from which these hybrid
SSPs originate. The figure shows energy profiles of three modes at the
particular subwavelength frequency, corresponding to a SSP/GW(c) mode at
$k_X=\pi/P$ (i), a GW mode of the unperforated structure (ii), and a pure SSP of a thick perforated metal film
surrounded by air (iii). Clearly, the hybrid SSPs
have the highest confinement and the highest concentration of
energy inside the dielectric film: $15\%$ ($70\%$) of the SSP/GW
(GW) mode resides outside of the dielectric slab.

Hybridization between FP modes and SSPs produce leaky hybrid
SSPs above the light line as shown in Fig.~\ref{Fig:DD} by the
progression from FP to SSP/FP modes. These hybrid {\it leaky} SSPs
have a similar three-dimensional magnetic field profile to
that of the {\it guided} SSP/GW modes as shown in the inset (a) to
Fig.~\ref{Fig:FieldProfiles}. Because fields are
concentrated inside the perforation, spectral properties and the
lifetime of the SSP/FP modes are strongly affected by the holes'
size $L_{X,Y}$. Their transverse confinement also significantly exceeds
that of FP modes as indicated by Q-factor shown by color in Fig.~\ref{Fig:DD}. The SSP/FP mode lifetime $Q/\omega$ and field confinment both increase with $L_{X,Y}$. The corresponding energy density profiles are similar to those plotted for the confined modes in Fig.~\ref{Fig:FieldProfiles}. Note that,
in addition to the SSP/FP modes, there is another leaky hybrid SSP
labeled SSP/GW(j) above the light line whose field distribution
(shown in Fig.~\ref{Fig:FieldProfiles} inset (b)) is centered outside of
the hole. Therefore, the spectral position of the leaky SSP/GW modes
is not strongly affected by the holes' dimensions. Because SSP/FP
and SSP/GW modes are leaky above the light line, they can be
experimentally observed using conventional infrared spectroscopy
as presented below.

\begin{figure}[ht] \centering
    \includegraphics[clip=true,scale=0.42]{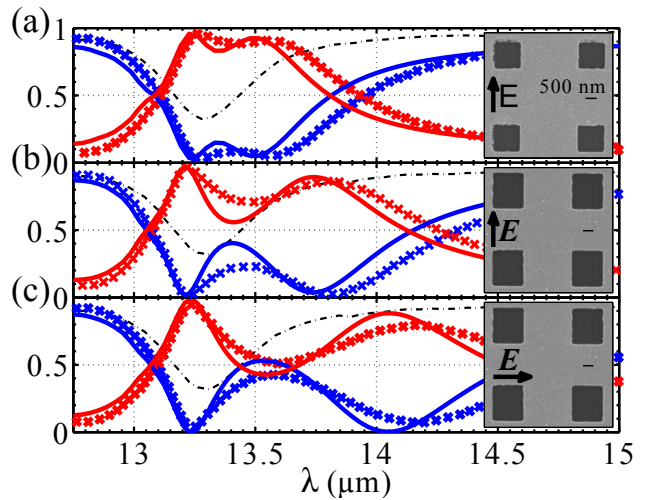}
    \caption{(color online). Infrared spectroscopy of hybrid spoof
      surface plasmons.  Reflection (blue) and absorption (red) curves are
      obtained from experimental measurements (crosses) and finite-element
      simulations (lines) for different hole dimensions \LX$\times$\LY{} and
      incident light polarizations. (a) \LX=\LY=1.2\um{},
      polarization-independent; (b) \LX=1.45\um{}, \LY=1.75\um{},
      $y$-polarized; (c) \LX=1.45\um{}, \LY=1.75\um{}, $x$-polarized. Black
      dotted curves: reflection from unperforated Au film/SiC
      structure. Periodicity $P$=$3.84$\um, SiC (Au) film
      thickness $d$=$570$\nm ($d_{\rm Au}$=$100$\nm.)}
\label{Fig:Expr}
\end{figure}

{\em Experimental results.} A SiC film of thickness $d$=$570$\nm{} was hetero-epitaxially
grown on a Si(100) substrate wafer, and an air-bridged membrane
with area ($500$\um)$^2$ was produced after KOH back-etching of
the Si substrate. A $100$\nm{} Au film was thermally deposited on
the newly exposed side of the SiC. Metal perforations were
produced using a focused ion beam over a ($100$\um)$^2$ area.
Optical characterization (reflection, transmission) was performed
using a Thermo Scientific Continuum microscope coupled to a
Nicolet 6700 FTIR Spectrometer. Incident radiation was linearly
polarized using a wire grid polarizer. In all measurements,
transmission was less than $10\%$. Therefore, we will focus on
reflectivity $R(\lambda)$ and absorptivity $A(\lambda)$ spectra
for the rest of this Letter.
First, we measured $R(\lambda)$ prior to perforating the metal film which is shown by dot-dashed lines in Fig.~\ref{Fig:Expr}.
The broad reflection dip at $\lambda_{\rm FP}$=$13.3$\um{} corresponds to
coupling to FP mode.

Next, we measured $R(\lambda)$ and $A(\lambda)$ for the perforated metal structures and plotted them alongside the results of the finite-elements COMSOL simulations in Fig.~\ref{Fig:Expr}. Two hole sizes were used as perforations: small square holes
(Fig.~\ref{Fig:Expr}(a)) and larger rectangular holes (Fig.~\ref{Fig:Expr}(b,c)).
As predicted by the theory of hybrid SSPs, reflectivity
curves reveal two reflection minima, which also correspond to
absorption maxima. Thus, the metal perforation causes $R(\lambda)$ to drop to zero and absorption $A(\lambda)$ to reach nearly unity. The physical reason for this enhanced absorptivity is resonant coupling to leaky SSP/FP (long-wavelength peak) and SSP/GW (short-wavelength peak) modes.

Experimental results confirm that the spectral position of the
SSP/GW dip ($\lambda$=$13.25$\um) is almost independent of the
size/shape of the holes, while that of SSP/FP dip is very
sensitive to hole size and, for rectangular holes, is
polarization-sensitive. The SSP/FP absorption peak
(e.g., at $\lambda$=$14.20$\um{} for light polarized along the
short dimension of the large rectangular hole) is especially remarkable
because it occurs far from the FP resonance in the spectral region
where SiC is not very absorptive: $n$=$4.6$+$0.04i$. Without
perforation, the round-trip absorptivity of the $d$=$\lambda/25$
SiC film is only $8\pi\,{\rm Im}(n)\,d/\lambda$=$4\%$. This absorption enhancement is caused by the excitation of the high-$Q$ SSP/FP mode which strongly traps and enhances the electromagnetic field inside the weakly-absorbing film, turning it into a ``perfect'' absorber.

\begin{figure}[ht]
\includegraphics[scale=0.48]{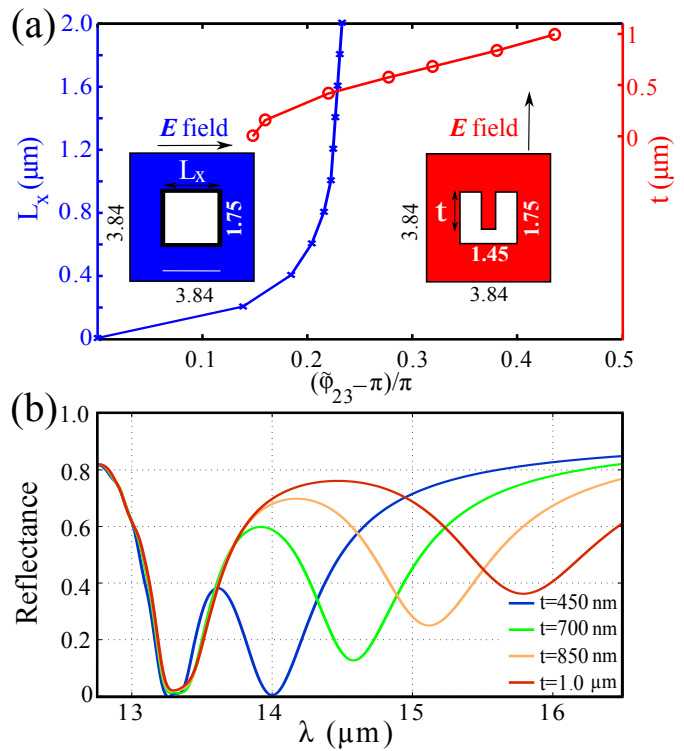}
  \caption{(color online). (a) Reflection phase shift
  $\tilde{\phi}_{23}$ from the
  metal film with holes at the reflectivity
  minima for (i) rectangular holes as the function of the holes' width
  $L_X$ (crosses), and (ii) U-shaped holes as the function of the
  tongue's length $t$ (circles). (b) Family of reflection
  spectra (different values of $t$ (nm)) for the U-shaped holes.}
  \label{Fig:HorseShoe}
\end{figure}

Another degree of freedom to confine electromagnetic energy is to
use more sophisticated hole shapes that exhibit site (or shape)
resonances~\cite{Sambles_PRL,Maier}. For example, a strong confinment even in a thinner heterostructures can be
accomplished by employing relatively simple (and feasible for fabrication) U-shaped holes~\cite{Rockstuhl} shown in
Fig.~\ref{Fig:HorseShoe}(a). For instance, in the case of the SSP/FP resonance, use of U-shaped holes causes even more dramatic
red-shifting which facilitates trapping of electromagnetic fields by a subwavelength thick heterostructure. 
A family of reflectivity spectra shown in Fig.~\ref{Fig:HorseShoe}(b) correspond to
U-shapes with increasing metal ``tongue'' length $t$ (see inset to
Fig.~\ref{Fig:HorseShoe}(a)). The figure indicates that, counter-intuitively,
the spectral position of the reflection dip red-shifts even as the
remaining hole area decreases with increasing $t$. This effect
is due to the LC-like resonances of the U-shape
occurring for electric field polarized
parallel to the tongue~\cite{Rockstuhl}. This resonance manifests
itself in the increased effective reflection phase
$\tilde{\phi}_{23}$. $\tilde{\phi}_{23}$ obtained for both rectangular and U-shaped holes at the
respective reflectivity dips is shown by crosses and circles in
Fig.~\ref{Fig:HorseShoe}(a) as a function of $L_X$ and $t$, respectively. The U-shape's
parameter $t$ has a more pronounced effect on $\tilde{\phi}_{23}$
(and, by extension, on the spectral position of the reflection
dip) than the overall size $L_X$ of the rectangular hole. In the
latter case, the reflection phase $\tilde{\phi}_{23}$ saturates
around the value of 1.25$\pi$, while in the case of the U-shaped
holes, $\tilde{\phi}_{23}$ approaches $3\pi/2$, with high value of
$d \tilde{\phi}_{23}/dt$ indicating strong sensitivity to the
geometry of the U-shape.

In conclusion, we have predicted and experimentally demonstrated
that a heterostructure comprised of a periodically perforated
metal film coated with high-index dielectric layer supports
strongly confined hybrid spoof surface plasmons. This
extraordinary confinement is achieved through hybridization
between two weakly confined sets of modes: SSPs supported by the
perforated metal film and guided modes supported by the dielectric
layer. Hybrid SSPs retain tunability due to strong sensitivity to the holes'
size and shape. As an experimental proof of principle, we have
demonstrated that hybrid SSPs can turn weakly-absorbing ultra-thin
semiconductor films into spectrally-tunable ``perfect'' absorbers.
Hybrid SSPs may find practical applications in multi-spectral
infrared imaging, thermophotovoltaics, on-chip light sources, and
infrared detectors.

This work was supported by NSF Grants ECCS-0709323 and
CMMI-0928664, and the AFOSR MURI Grants FA9550-06-1-0279 and
FA9550-08-1-0394.


\begin{thebibliography}{99}

\bibitem{barnes_nature03} W.~L.~Barnes, A.~Dereux, and
T.~W.~Ebbesen, Nature {\bf 424}, 824 (2003).

\bibitem{Raether_book} H. Raether, {\it Surface Plasmons on Smooth and
Rough Surfaces and on Gratings}, (Springer, Berlin, 1988).

\bibitem{halas_natphot07} S.~Lal, S.~Link, and N.~J.~Halas, Nature Phot.~{\bf 1}, 641 (2007).

\bibitem{Atwater} H. A. Atwater and A. Polman, Nature Mat. {\bf 9}, 205
(2010).

\bibitem{bozh_nature06} S.~I.~Bozhevolnyi, V.~S.~Volkov, E.~Devaux,
J.-Y.~Laluet, and T.~W.~Ebbesen, Nature {\bf 440}, 508 (2006).

\bibitem{Pendry_sc} J. B. Pendry, L. Mart{\' \i}n-Moreno, and
F. J. Garc{\' \i}a-Vidal, Science {\bf 305}, 847 (2004).

\bibitem{Abajo_PRL} F. J. Garc{\' \i}a de Abajo and J. J. S{\' a}enz, Phys. Rev. Lett. {\bf 95}, 233901 (2005).

\bibitem{Abajo_OptExp} X. M. Bendana and F. J. Garc{\' \i}a de Abajo, Opt. Exp. {\bf 17}, 18826 (2009).

\bibitem{Abajo_Review} F. J. Garc{\' \i}a de Abajo,
Rev. Mod. Phys. {\bf 79}, 1267 (2007).

\bibitem{Sambles_PRL} M. J. Lockyear, A. P. Hibbins, and J. R. Sambles, Phys. Rev. Lett. {\bf 102}, 073901 (2009).

\bibitem{Maier} M. Navarro-C{\' \i}a, M. Beruete, S. Agrafiotis,
F. Falcone, M. Sorolla, and S. A. Maier,
Opt. Exp. {\bf 17}, 18184 (2009).

\bibitem{Ozawa_JPC91} M.~Osawa and M.~Ikeda, J.~Phys.~Chem. {\bf
95}, 9914 (1991).

\bibitem{Halas_CPL08} J.~Kundu, F.~Le, P.~Norlander, and
N.~J.~Halas, Chem.~Phys.~Lett.~{\bf 119}, 452 (2008).


\bibitem{painter_optexp10} J.~Rosenberg, R.~V.~Shenoi, S.~Krishna, and
O.~Painter, Opt.~Exp.~{\bf 18,} 3672 (2010).

\bibitem{brongersma_optexp08} A.~C.~Hryciw, Y.~C.~Jun, and
M.~L.~Brongersma, Opt.~Exp.~{\bf 17,} 185 (2008).

\bibitem{shawnlin_NL10} C-C. Chang, Y. D. Sharma, Y-S. Kim, J. A. Bur, R. V. Shenoi,
S. Krishna, D. Huang, and S-Y. Lin, Nano Letters {\bf 10,} 1704
(2010).

\bibitem{neuner_jpc10} B.~Neuner III, D.~Korobkin, C.~Fietz,
D.~Carole, G.~Ferro, and G.~Shvets, J.~Phys.~Chem. C {\bf 114},
7489 (2010).

\bibitem{born_and_wolf} M. Born and E. Wolf, {\it Principles of
Optics} (Pergamon Press, New York, 1980), 6th ed.



\bibitem{Martin-Moreno_PRL} L. Mart{\' \i}n-Moreno, F. J. Garc{\'
\i}a-Vidal, H. J. Lezec, K. M. Pellerin, T. Thio, J. B. Pendry,
and T. W. Ebbesen, Phys. Rev. Lett. {\bf 86}, 1114 (2001).

\bibitem{Rockstuhl} C. Rockstuhl, T. Zentgraf, T. P. Meyrath,
  H. Giessen, and F. Lederer, Opt. Exp. {\bf 16}, 2080 (2008).


\end{thebibliography}
\end{document}